# A Novel Isomer of Volleyballene Sc20C60

Peng-Bo Liu,[a] Jing-Jing Guo,[a] Hui-Yan Zhao,*,[a] Hong-Man Ma,[a] Jing Wang,[a] and Ying Liu*,[a]

[a] Dr. P. B. Liu, Dr. J. J. Guo, Dr. H. Y. Zhao, Dr. H. M. Ma, Prof. J. Wang, Prof. Y. Liu
Department of Physics and Hebei Advanced Thin Film Laboratory, Hebei Normal University
Shijiazhuang, 050024 Hebei, China
E-mail: hyzhao@hebtu.edu.cn, yliu@hebtu.edu.cn
Homepage: https://www.researchgate.net/profile/Ying_Liu86

Supporting information for this article is given via a link at the end of the document.

**Abstract:** The Stone-Wales defect is a well-known and significant defective structure in carbon materials, impacting their mechanical, chemical, and electronic properties. Recently, a novel metal-carbon nanomaterial named Volleyballene has been discovered, characterized by a C-C bond bridging two carbon pentagons. Using first-principles calculations, a stable Stone-Wales-defective counterpart of Volleyballene, exhibiting $T_h$ symmetry, has been proposed by rotating the C-C bond by 90°. Although its binding energy per atom is slightly higher than Volleyballene ($\Delta E_b = 0.009$ eV/atom), implying marginally lower structural stability, it can maintain its bond structure until the effective temperature reaches about 1538.56 K, indicating greater thermodynamic stability. Additionally, its highest vibration frequency is 1346.2 cm-1, indicating strong chemical bond strength. A theoretical analysis of the $Sc_{20}C_{60}+Sc_{20}C_{60}$ binary systems highlights that the stable building block may be applied in potential nano-assembly.

## Introduction

Since the discovery of $C_{60}$, novel clusters have attracted scientists' interest. The study of Metallo-Carbohedrenes (Met-Cars) showed that $Ti_8C_{12}^+$ has high mass distribution abundance in the reactions of titanium with $CH_4$ and $C_2H_2$. [1] Similar clusters $M_8C_{12}$ ($M$ = Zr, HF, V, Nb, Mo, Cr, etc) were found, [2,3] and $Ti_{14}C_{13}^+$ and $V_{14}C_{13}^+$ of other components also showed similar stability to $Ti_8C_{12}$. [4-6] In recent years, Majumder's group have studied its hydrogen storage capacity, [7] and Sofo et al. have studied the stability of Met-Cars after metal element replacement. [8] Wang et al. designed $Ti_2C$ sheet and researched its stability, mechanical and thermal properties. [9] Recently, Jena et al. constructed one-dimensional nanowires using $Ti_9C_{13}$ clusters as well as those based on planar-tetra-coordinate carbon-containing (ptC) $Ti_8C_{12}$ Met-Car and TiC clusters with bulk cubic structure when designing high-capacity anodes for lithium-ion batteries (LIB). [10] Their energy stability, thermal stability and potential as LIB anode were studied using density functional theory and molecular dynamics. [10] Agúndez et al. have researched the chemical equilibrium in the atmosphere of the asymptotic giant branch (AGB) star. [11] They investigated $Ti_xC_y$ ($x$ = 1- 4; $y$ = 1- 4) clusters for the first time, and selected larger clusters up to $Ti_{13}C_{22}$. The chemical equilibrium prediction showed that large titanium carbon clusters, such as $Ti_8C_{12}$ and $Ti_{13}C_{22}$, were existed in the atmosphere of AGB star.

More recently, our research group proposed and identified volleyball-shaped metal-carbon nanomaterial $Sc_{20}C_{60}$, and named it Volleyballene (here which is called Volleyballene-I). It is constructed from octagon rings of scandium atoms and pentagon rings of carbon atoms, and the structure is energetically and dynamically stable. [12] The Volleyballene molecule may be able to accommodate other atoms or molecules for the purpose of studying fundamental chemistry. [13,14] Tlahuice-Flores et al. have also researched its stability and hydrogen storage characteristics. [15-17]

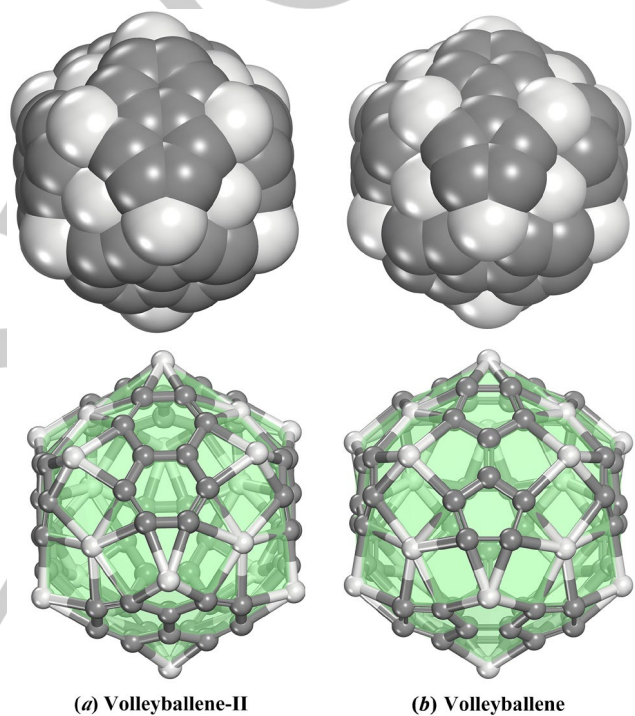

**Figure 1.** The comparison of the geometric structures of Volleyballene-II (*a*) and the original Volleyballene (*b*). The CPK model above and the ball and stick model down below. The silver white sphere represents Sc atom, and the gray sphere represents C atoms.

The well-known Stone-Wales (SW) defect is a prototypical and ubiquitous intrinsic point defect of carbon nanomaterials. [18-20] It has multiple influence on their physical and chemical properties, and is formed by rotating a C-C bond by an angle of 90° (the so-called Stone-Wales transformation). In this manuscript, another Volleyballene $Sc_{20}C_{60}$ variant, Volleyballene-II, was formed by a SW-like transformation of a C-C bond bridging two five-membered rings in the $Sc_8C_{10}$ sub-unit of Volleyballene-I.

## Results and Discussion

### 3.1 Geometry Structure

As shown in Figure 1, just like Volleyballene-I, [11] Volleyballene-II still consists of 20 scandium atoms and 60 carbon atoms, and still consists of six $Sc_8C_{10}$ subunits which joined together according to the pattern of volleyball. Eight Sc atoms locate on the edge of the subunit. According to the difference of atomic coordination number and adjacent atoms, the C atoms are divided to three types ($C^I$, $C^{II}$ and $C^{III}$), and the Sc atoms are divided to two types ($Sc^I$ and $Sc^{II}$) shown from Table 1. Each $Sc^I$ atom is shared by three subunits, and each $Sc^{II}$ atom is divided by two subunits. In the $Sc_8C_{10}$ subunit of Volleyballene-II, 10 C atoms are composed of two six-membered carbon rings with sharing a C-C bond, while in the $Sc_8C_{10}$ subunit of Volleyballene-I, two five-membered carbon rings are bridged by a C-C bond. Volleyballene-II could be formed from Volleyballene-I by C-C bond-rotation in the $Sc_8C_{10}$ sub-unit.

Table 1 gives the structural information of Volleyballene-II after energy minimization. Some typical bond lengths and interatomic distances are listed in Table 1. The distance between $Sc^I$ and $Sc^{II}$ atoms is up to 3.160 Å, indicating that adjacent Sc atoms do not form bonds.

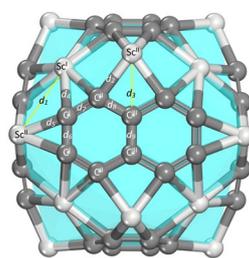

**Table 1.** Some typical bond lengths (━) and interatomic distances (↔) for Volleyballene-II. Here, $Sc^I \leftrightarrow Sc^{II}$ ($d_1$) and $Sc^{II} \leftrightarrow C^{III}$ ($d_3$) represent the inter-atomic distance (represented by yellow lines in the figure).

| | | |
|---|---|---|
| $Sc^I \leftrightarrow Sc^{II}$ | $d_1$ | 3.160 |
| $Sc^{II} \boldsymbol{-} C^{II}$ | $d_2$ | 2.143 |
| $Sc^{II} \leftrightarrow C^{III}$ | $d_3$ | 2.414 |
| $Sc^I \boldsymbol{-} C^I$ | $d_4$ | 2.346 |
| $Sc^{II} \boldsymbol{-} C^I$ | $d_5$ | 2.143 |
| $C^I \boldsymbol{-} C^I$ | $d_6$ | 1.430 |
| $C^I \boldsymbol{-} C^{II}$ | $d_7$ | 1.412 |
| $C^{II} \boldsymbol{-} C^{III}$ | $d_8$ | 1.455 |
| $C^{III} \boldsymbol{-} C^{III}$ | $d_9$ | 1.467 |

**Table 2.** Comparison of Volleyballene-I and Volleyballene-II at the PBE levels. The data include symmetry (sym.), average C-C bond length ($d_{C-C}$), average Sc-C bond length ($d_{Sc-C}$), average distance between Sc atoms ($d_{Sc-Sc}$), relative binding energy per atom ($\Delta E_b$), HOMO-LUMO gap ($E_g$), HOMO energy level and average charge transfer from the Sc atom ($Q_{Sc}$). The units of distance, energy, and charge are Å, eV/atom, and e, respectively.

| Isomers | Sym. | $d_{C-C}$ | $d_{Sc-C}$ | $d_{Sc-Sc}$ | $\Delta E_b$ | $E_g$ | $E_{HOMO}$ | $Q_{Sc}$ |
|---|---|---|---|---|---|---|---|---|
| Volleyballene-I | $T_h$ | 1.447 | 2.242 | 3.220 | 0.000 | 1.463 | -3.410 | 0.518 |
| Volleyballene-II | $T_h$ | 1.435 | 2.265 | 3.159 | 0.009 | 0.412 | -2.284 | 0.454 |

Table 2 shows some comparison results of the Volleyballene-I and Volleyballene-II at the PBE level. They all have $T_h$ symmetry, and the binding energy per atom of Volleyballene-II is slightly higher than that of Volleyballene-I. Compared with Volleyballene-I, the average distance of C-C and Sc-Sc in the Volleyballene-II is shorter, while the average bond length of Sc-C is longer. In the Volleyballene-I, the bond angle of $Sc^I$-$C^I$-$Sc^{II}$ is 93.330°, while in the Volleyballene-II, there are two types of bond angle of Sc-C-Sc. Furthermore, the bond angle of Sc-C-Sc is smaller than that of the Volleyballene-I, in which the bond angle of $Sc^I$-$C^{II}$-$Sc^{III}$ was 88.672° and that of $Sc^I$-$C^I$-$Sc^{II}$ was 89.397°. In addition, the diameters of the Volleyballene-I are about 9.862 Å-10.534 Å, and those of the Volleyballene-II are about 8.674 Å-10.698 Å. By the Stone-Wales transformation, the distance between the two opposite C atoms decreases in the $Sc_8C_{10}$ sub-unit, while the distance between the two opposite Sc atoms increases. Additionally, the atomic coordinates of the Volleyballene-I and Volleyballene-II were shown in Table S1 and Table S2.

### 3.2 Structural Stability

To test the thermodynamic stability of Volleyballene-II, snapshots of the NVT MD simulations at different times for each initial temperature are presented in Figure 2. The calculation results show that Volleyballene-II can maintain its chemical bond geometry well in the 5.0 *ps* molecular dynamics simulation at an initial temperature of 1500 K (with an effective temperature of around 1538.56 K). In contrast, for Volleyballene-I, this occurs at approximately 1000 K, highlighting its superior thermodynamic stability. Here the effective temperature is defined to the arithmetic mean of temperature values at all dynamics steps of a MD calculation. When the initial temperature simulated by NVT MD exceeds 1600 K, the structure of Volleyballene-II becomes unstable, and the geometric structure of Volleyballene-II was changed in various degrees. It can be concluded from Figure 2 that at 1600 K (the effective temperature is about 1640.84 K) and 1900 K (the effective temperature is about 1948.57 K), a six-membered carbon ($C_6$) ring was split into a five-membered carbon ring and a carbon atom (labeled in pink) that bridged with the five-membered carbon ring, and the bond structures of other six-membered carbon rings were not destroyed; While at 1700 K (the effective temperature is about 1743.39 K), the bond structure of two $C_6$ rings in one $Sc_8C_{10}$ sub-units was seriously damaged; In another $Sc_8C_{10}$ sub-units, a $C_6$ ring was split into a five-membered carbon ring and a carbon atom that bridges with the five-membered carbon ring; When the temperature rises to 1800 K (the effective temperature is about 1846.45 K), no significant structural "melting" occurred within 5.0 *ps*; Until the temperature goes to 2000 K (the effective temperature is about 2051.03 K), the cluster was partially melted and the bond structure was obviously distorted, but there were still some six-membered carbon rings and five-membered carbon rings. In conclusion, when the temperature exceeds 1600 K, although the bond structure of Volleyballene-II is irreversibly destroyed during the process of 1600 K NVT MD simulation, the six-membered carbon rings can still be partially maintained.

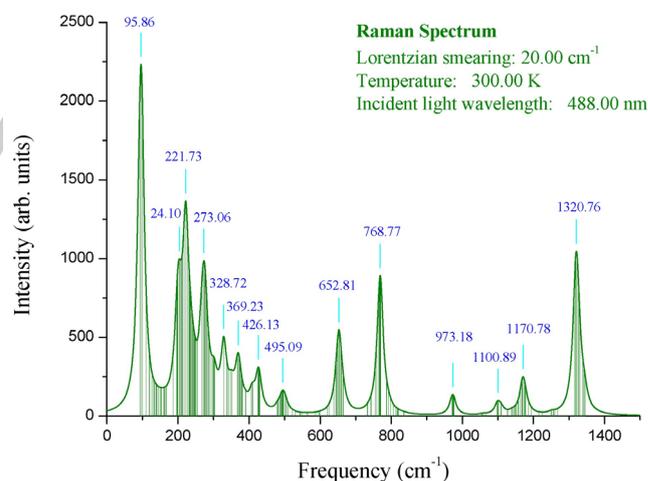

**Figure 3.** Raman spectrum for Volleyballene-II. A Lorentzian broadening of 20 cm⁻¹ is used. The labels give the frequency values corresponding to the peaks of the intensities.

The vibration frequency of Volleyballene-II and the vibration intensity of corresponding vibration modes were calculated. According to the calculation results, the minimum vibration frequency was at 93.3 cm⁻¹, and the maximum vibration frequency was at 1346.2 cm⁻¹. While for the Volleyballene-I, the vibration frequency ranges from 80.1 cm⁻¹ to 1389.5 cm⁻¹. There were no imaginary frequencies, which reflects the mechanical stability of Volleyballene-II. It means that the geometry is corresponding to the energy minimum point of the potential energy surface. The highest vibration frequency of 1346.2 cm⁻¹ also reflects the high strength of the chemical bonds of Volleyballene-II. In addition, under the conditions of the temperature of 300 K and the incident wavelength of 488.00 nm, the infrared activity and Raman intensity of Volleyballene-II were calculated. The calculation results were shown in Figure 3. A series of characteristic peaks of Raman intensity appear at 95.86, 221.73, 273.06, 768.77, and 1320.76 cm$^{-1}$, etc.

### 3.3 Electronic Structure

In order to further understand its stability and potential application, the electronic structure of Volleyballene-II was analyzed. Fig. 4(a) shows the electronic energy levels, and the partial density of states (PDOS) of Volleyballene-II. The calculation results show that the HOMO–LUMO gap of Volleyballene-II is 0.412 eV. Compared with Volleyballene-I, the Fermi level of the Volleyballene-II moves up, and the HOMO-LUMO gap becomes smaller. The spin-up and spin-down curves in PDOS are symmetric about the energy axis in Figure 4(a), indicating that the structure has no net spin magnetic moment. Hirshfeld analysis (Table S3) for atomic charge and spin also support that the total spin magnetic moment of Volleyballene-II is 0.00 $\mu_B$.

Figure 4 also shows electron density (b) and deformation electron density (c). Based on the deformation electron density analysis, it is evident that Sc atoms lose electrons, while C atoms exhibit both electron loss and gain, which is in conformity with the findings of the Hirshfeld analysis. Specifically, by the Stone-Wales transformation, the C in the Volleyballene-II can be divided into three types, named C$^I$, C$^{II}$ and C$^{III}$, with the average electron transfer of 0.020$e$, -0.166$e$ and -0.263$e$, respectively. And average charge transfer from each Sc atom to the nearby C atom is about 0.454$e$, as shown in Table 2. Besides, the accumulation of electrons is primarily seen within the C-C and Sc-C bonds, with no charge accumulation observed between the Sc atoms. This suggests the presence of a chemical bond between the C and Sc atoms, but no bond between the Sc atoms. This finding is consistent with the NBO results.

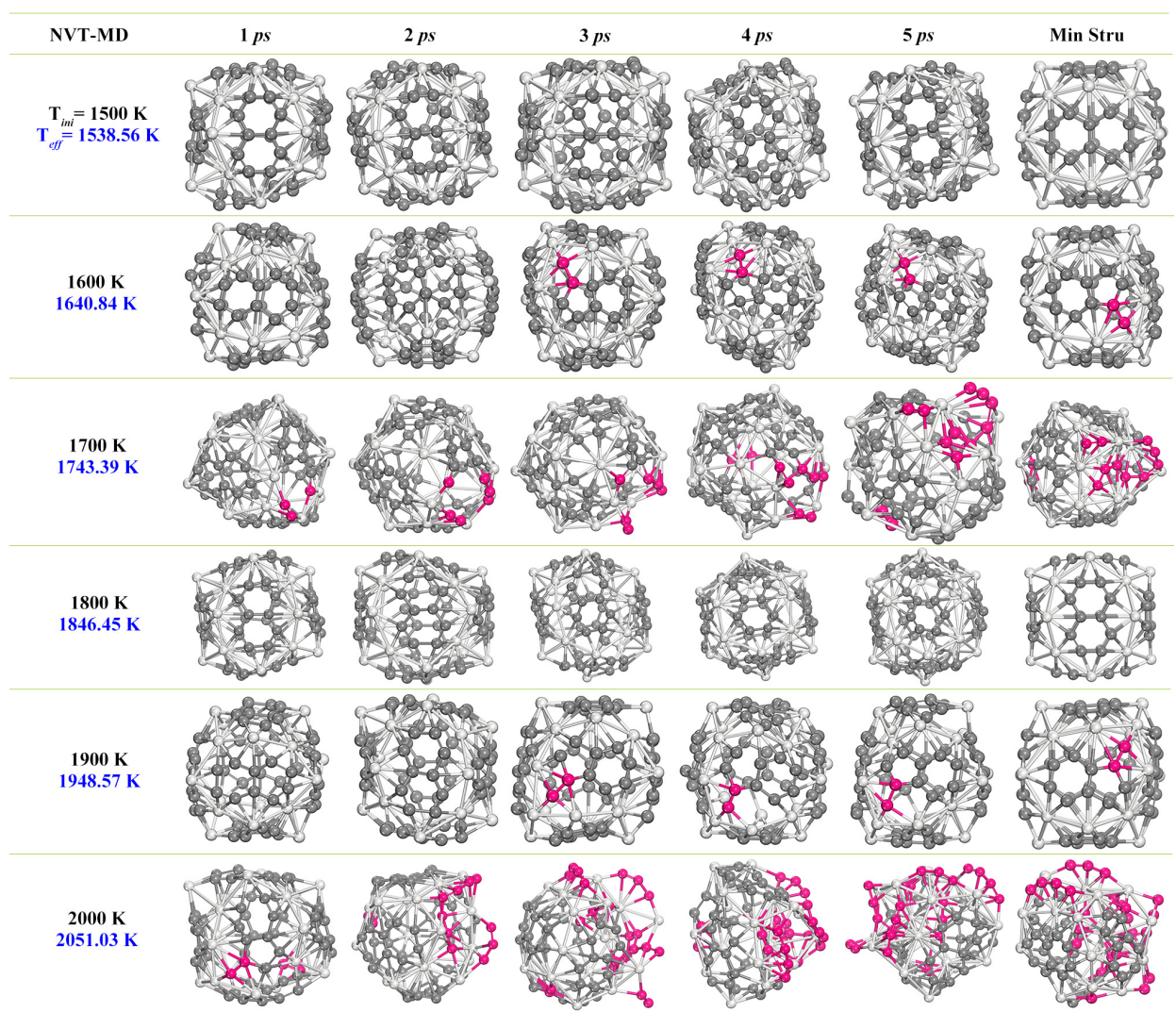

**Figure 2.** Snapshots for the NVT MD simulations at temperature 1500, 1600, 1700, 1800, 1900 and 2000 K at 1, 2, 3, 4, 5 $ps$. The blue temperature T$_{eff}$ corresponding to the initial temperature T$_{ini}$ on the leftmost side in the figure is the effective temperature value corresponding to NVT MD simulations within 5 $ps$. The rightmost row structure (Min Stru) in the figure is the structure at 5 $ps$ after energy minimization. In the figure, the carbon atoms with seriously damaged structure of the six-membered ring of carbon are marked in pink.

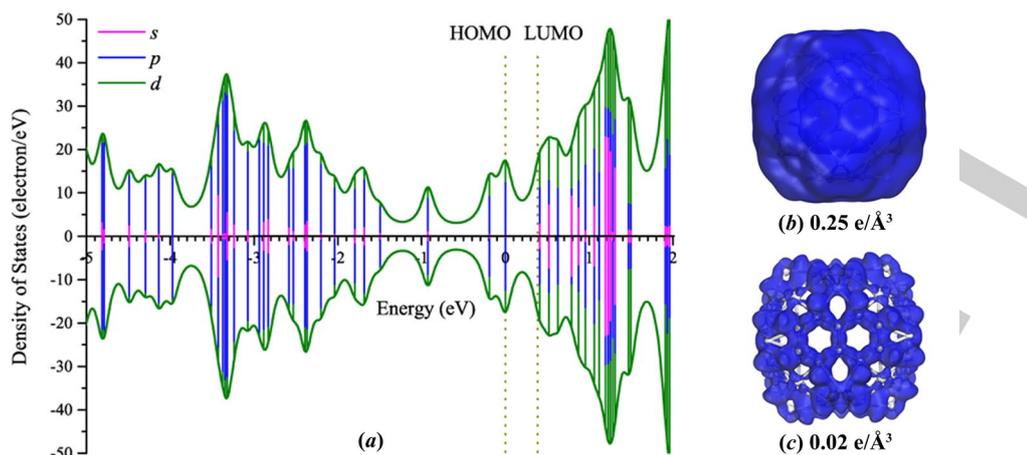

**Figure 4.** (*a*) Spin PDOS and HOMO, LUMO orbitals for Volleyballene-II at the PBE level. The zero of energy for the PDOS curve is taken to be the Fermi energy, and the length of each vertical line represents the corresponding relative amplitude for each electronic orbital. The coloured vertical lines in the PDOS envelope curve show the relative magnitudes of the electron states for several molecular orbitals: pink line for *s* orbitals; blue line for *p* orbitals, green line for *d* orbitals, and dark yellow dot lines for both HOMO and LUMO. The HOMO–LUMO gaps of Volleyballene-II is 0.412 eV. (*b - c*) The electron density (*b*) and deformation electron density (*c*) of Volleyballene-II. The iso-surface of electron density map (*b*) is set as 0.25 e/Å³, and the iso-surface of deformation electron density map (*c*) is set as 0.02 e/Å³. The blue part represents electron accumulation regions.

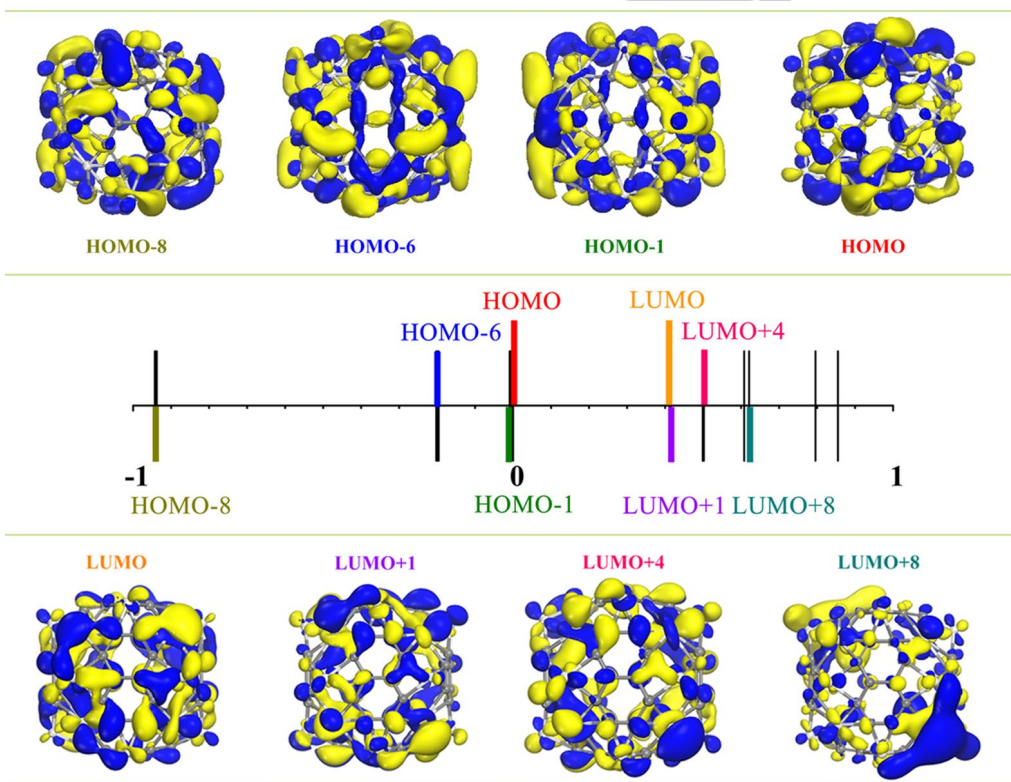

**Figure 5.** Selected frontier orbitals for Volleyballene-II. Center is the schematic diagram of orbital energy levels near the HOMO and LUMO orbitals and the inside color lines correspond the orbitals, i.e. dark yellow: HOMO-8, blue: HOMO-6, olive: HOMO-1, red: HOMO, orange: LUMO, violet: LUMO+1, pink: LUMO+4, and dark cyan: LUMO+8. The iso-surface is set to be 0.015 e/Å³.

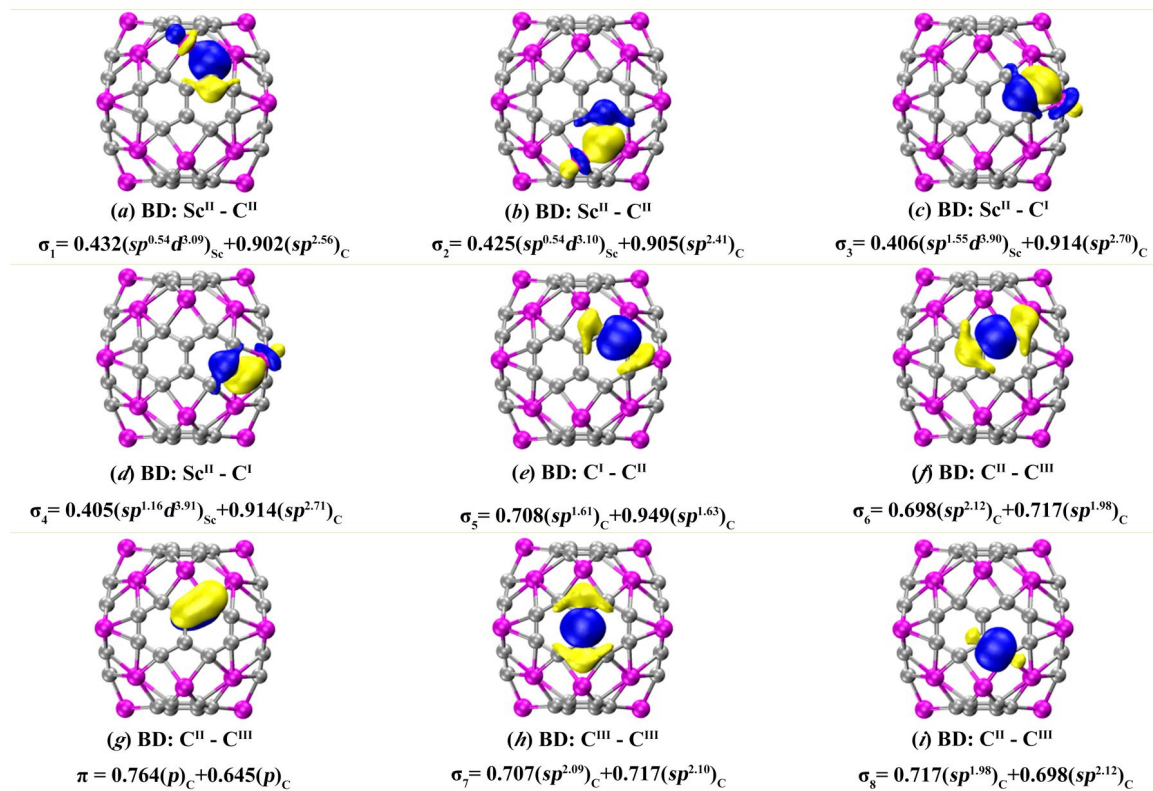

**Figure 6.** Some typical natural bond orbitals for the Volleyballene-II. The first lines list the type of atoms that make up the bond and the next lines summarize the natural atomic hybrids of this natural bond orbital.

The PDOS curve shows that there are obvious *p-d* hybridization characters in the HOMO, and other orbitals near HOMO shows the characters of *s-p-d* hybridization. Figure 5 shows some selected frontier orbitals of Volleyballene-II. At the $Sc^I$ atom location, the characteristics of $2d_z^2$-like orbital are present at HOMO-6, LUMO, LUMO+1, and LUMO+8 energy levels. Different types of *d*-like orbitals ($d_{x^2-y^2}$, $d_{xy}$, $d_{yz}$, $d_{xz}$) are shown for Sc atoms at HOMO-8, HOMO-1, HOMO, and LUMO+4. Also there are some significant characteristics of π orbital at the C atoms in Figure 5.

Figure 6 shows some typical natural bond orbitals for the Volleyballene-II. For each Sc atom in the $Sc_8C_{10}$ sub-unit, a $d_z^2$-like orbital forms an σ bond with a neighboring carbon atom, as plotted in Figure 6*a*-6*d*. In carbon hexagon structure, there are two-center two-electron σ bonds (Figure 6*e*-6*f*) and π bonds (Figure 6*g*). For all σ(C-C) bonds, it can be found that major contribution comes from $sp^2$-like hybridization of C atom. The hybrids of $sp^{1.63}$, $sp^{1.98}$, $sp^{2.10}$ and $sp^{2.12}$ are all distorted $sp^2$ hybridizations, as plotted in Figure 6*e*, Figure 6*f*, Figure 6*h* and Figure 6*i*. For C atom, the π bond is formed by the hybridization of *p* orbitals as shown in Figure 6*g*.

### 3.4 Binary Systems

Nanocluster assembled materials have attracted great interest for designing nanostructured materials with unique properties. Clusters, as the building blocks of nanocluster assembled materials, should be stable enough to resist dissociation and aggregation during preparation and manipulation. In order to further examine the stability of Volleyballene-II and whether cluster aggregation will occur, four representative configurations of Volleyballene-II binary systems were discussed, and the total energy of each configuration with the change of the distance between two Volleyballene-II sub-units was shown. In order to judge the independence of the Volleyballene-II, we analyzed the electronic structure of each configuration at the equilibrium distance, and the calculation results are shown in Figure 7.

In the configurations (a) and (b), the $Sc_8C_{10}$ sub-units in two Volleyballene-IIs are relative (see Figure 7). In the configuration (a), two $Sc_8C_{10}$ sub-units are mirror symmetric, and the atoms of the same type are directly opposite. The relative position of the two clusters in configuration (b) is achieved by rotating the cluster on the right side of configuration (a) 90° to the left or right around the center-line that connects the two clusters. In the configurations (c) and (d), the curved triangles of $Sc_4C_6$ in two Volleyballene-IIs are opposite (see Figure 7). In the configuration (c), two $Sc_4C_6$ units are symmetrical about the vertical plane of the central line of the two clusters, and the atoms of the same type are directly opposite. In configuration (d), the relative position of the two clusters is achieved by rotating the right cluster in configuration (c) 30° to the left or right around the centerline that connects the two clusters.

To calculate the change in total energy of the binary system as a function of the distance between two Volleyballene-II clusters, we use the increment $d = \Delta x$ of the *x* component of the minimum atomic coordinate of two Volleyballene-II clusters on the *x* axis (transverse) to represent the distance between two Volleyballene-II clusters. When calculating the total energy, we fixed the positions of the atoms in each Volleyballene-II cluster far away from the center of each cluster. In the calculation, T-S schemes was used to calculate long range dispersion interactions.

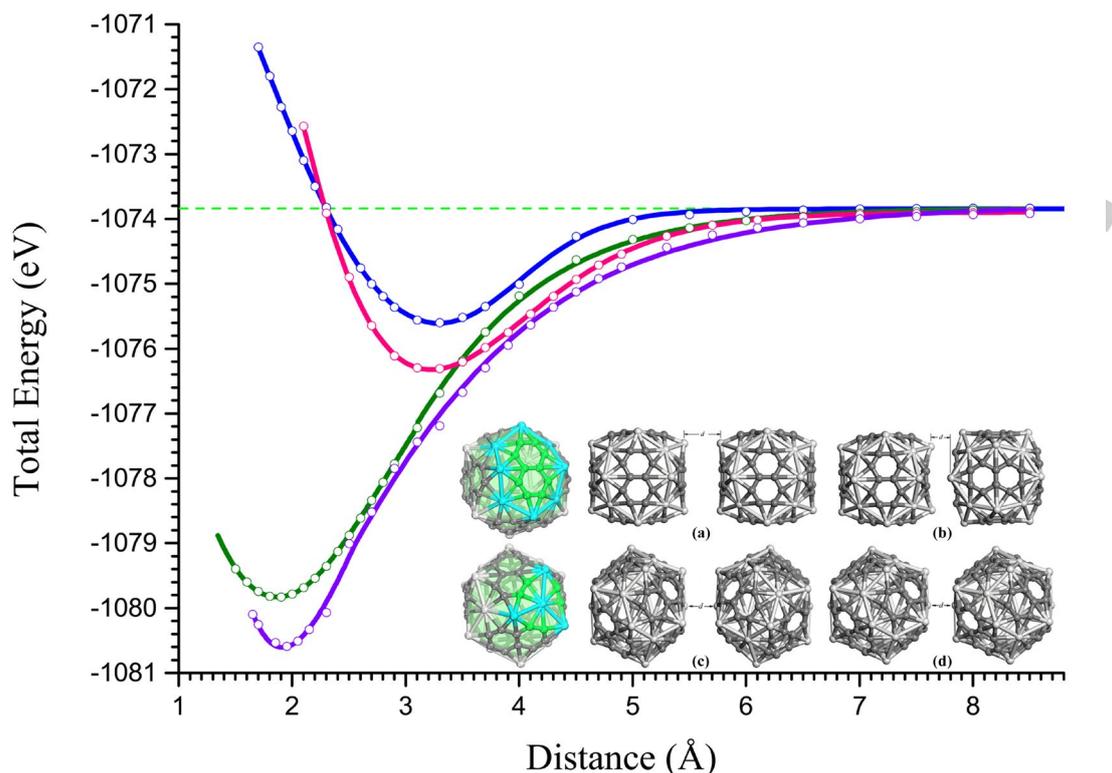

**Figure 7.** The total energy of four typical Volleyballene-II binary systems varies with the minimum distance between two Volleyballene-IIs in the system. The lower right in the figure is the configuration of four representative Volleyballene-II binary systems. The adjacent two faces in binary systems are corresponding to the structures with green and cyan highlighted atoms. In each configuration, the minimum distance between two Volleyballene-IIs in the initial state is marked with $d$. The four total energy curves in the figure are represented by blue (a), olive (b), pink (c) and violet (d). The green dash line indicates the total energy when the distance between two Volleyballene-IIs is $d \to \infty$. The cluster spacing at the lowest point of the total energy curve is expressed by $d_0$.

Figure 7 shows the change of total energy of Volleyballene-II binary system with cluster spacing in four configurations. It can be seen from the curve that with the increase of interatomic distance, two Volleyballene-II clusters in the four configurations show a change from mutual exclusion to mutual attraction. When the distance between two clusters exceeds 9 Å, the interaction energy tends to zero, and the total energy of Volleyballene-II binary system tends to be twice that of a single Volleyballene-II. The calculated data of the total energy of the Volleyballene-II binary system varying with the cluster spacing is obtained by non-linear curve fit. The equilibrium distance $d_0$ (the distance corresponding to the minimum of the total energy) of the two clusters are: (a) $d_0 = 3.263$ Å, (b) $d_0 = 1.870$ Å, (c) $d_0 = 3.215$ Å, (d) $d_0 = 1.941$ Å, respectively.

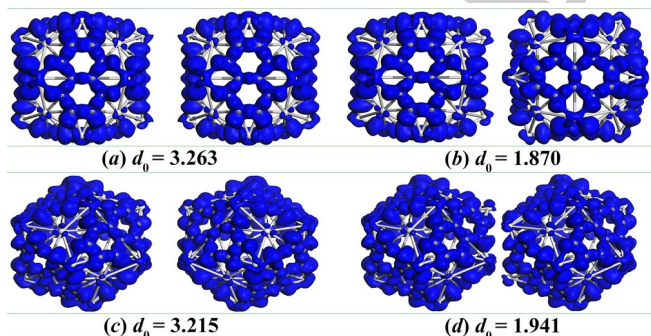

**Figure 8.** Deformation charge density of four typical Volleyballene-II systems at equilibrium position $d_0$ (Å). The iso-surface is set to 0.05 e/Å³. The blue part represents charge accumulation regions.

Figure 8 shows the deformation electron densities of the four configurations of the Volleyballene-II binary system at the equilibrium spacing. The iso-surface of deformation electron densities is taken to be

0.05 e/Å³. The results of deformation electron densities show that there is no obvious bonding feature were detected between the two clusters at the equilibrium position, and each Volleyballene-II still maintains relative independence and structural integrity.

## Conclusion

In this work, we proposed the Volleyballene-II, a Stone-Wales-defective counterpart of Volleyballene $Sc_{20}C_{60}$. The molecular dynamics calculations and vibration frequency calculations proved that its structure is stable. Its binding energy per atom is slightly higher than that of the original volleyballene ($\Delta E_b = 0.009$ eV/atom), which can be regarded as an energy nearly degenerate isomer. Molecular dynamics calculations show that it can maintain its bond structure at higher effective temperature, indicating its better thermodynamic stability. The research of the Volleyballene-II binary system also shows that Volleyballene-II can exist stably as a "building element" for the construction of nanomaterials. Similar to Volleyballene-I, Volleyballene-II is a large, hollow cluster whose properties can be regulated by embedded atoms or small molecules, which makes it possible to explore new super atomic structural chemistry.

## Computation Methods

Our calculations were performed within the framework of spin-polarized density-functional theory (DFT). Perdew–Burke–Ernzerhof (PBE) [21], the Perdew-Wang 1991 (PW91) [22], and the Becke in conjunction with the Lee-Yang-Parr (BLYP) [23] levels with the generalized gradient approximation (GGA) were all adopted to conduct symmetry-unconstrained geometry optimizations. The structure was allowed to relax freely without any consideration of symmetry. A double numerical plus polarization (DNP-4.4) basis set[24] with all electron calculation method was employed. Here, T-S dispersion correction were put forward by Tkatchenko and Scheffler to consider the long-range dispersion interactions.[25] The convergence thresholds for the geometry optimization were set as $1 \times 10^{-5}$ Ha for energy, $2 \times 10^{-5}$ Ha/Å for forces, and $5 \times 10^{-3}$ Å for maximum displacement. The coresponding calcualtion results were listed in Table S4. It can be clearly observed that the results of different exchange-correlation functions (PBE, PW91, and BLYP) are qualitatively the same. What's more, the GGA-PBE functional with DNP-4.4 basis set and T-S dispersion correction has been proved to be more reliable DFT method to model the structure.[26-27]

Subsequently, *ab initio* molecular dynamics (MD) simulations in canonical ensembles (NVT: constant-temperature, constant-volume ensemble) were performed with a Gaussian thermostat.[28] The total simulation time and time step were set to 5 *ps* and 1 *fs*, respectively. To ensure the adequacy of the simulation time of 5 *ps*, relevant test calculations were performed. Firstly, the potential energy profile in MD simulation at 1500 K was plotted and shown in Figure S1. The potential energies observed at different time intervals, as shown in Figure S1, eventually reach an equilibrium state and exhibit an oscillation pattern centered around a nearly constant potential energy value. Secondly, a long total simulation time of 10 *ps* for MD simulations was conducted to explore whether simulation time of 5 *ps* is long enough to characterize the isomer stability. It can be found that even undergo 10 *ps* MD simulations, final structure of the new isomer not changed much compared with the final configuration of 5 *ps* NVT MD simulation. Our calculations above were applied out in the DMol³ package. [24]

To understand the chemical bonding, the natural bond orbital (NBO) analysis was performed using Gaussian 09 package.[29] Multiwfn software[30] and Visual Molecular Dynamics (VMD) software[31] are used to analyze and visualize calculation results.

## Acknowledgements

This work is supported by the National Natural Science Foundation of China (Grant No. 12174084), the Natural Science Foundation of Hebei Province (Grant No. A2021205024), and the Key Program of Scientific and Technological Foundation of Hebei Province (Grant No. ZD2021065).

**Keywords:** Carbon nanomaterials • Molecular Geometry • Stability • Stone-Wales defect • Volleyballene

**Entry for the Table of Contents**

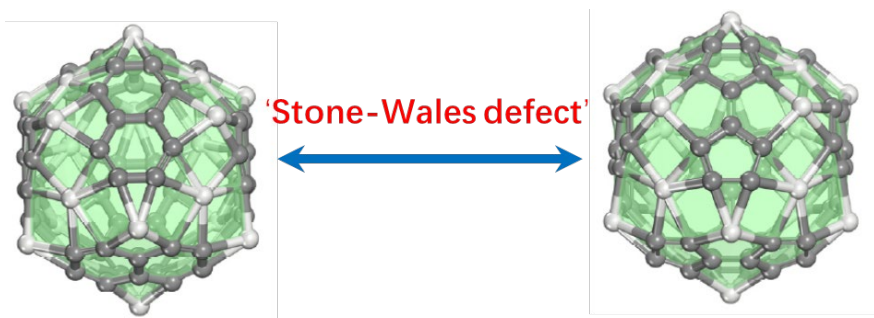